\documentstyle[aps,preprint]{revtex}
\newcommand{\bea}{\begin{eqnarray}}
\newcommand{\beq}{\begin{equation}}
\newcommand{\eea}{\end{eqnarray}}
\newcommand{\eeq}{\end{equation}}
\begin{document}
\title{Semiclassical analysis of a two-electron quantum 
dot in a magnetic field: dimensional phenomena}
\author{R.G. Nazmitdinov $^{1,2}$, N. S. Simonovi\'c $^1$,
and Jan M. Rost $^1$}
\address{
$^1$ Max-Planck-Institut f\"ur Physik komplexer
Systeme, D-01187 Dresden, Germany\\
$^2$ Bogoliubov Laboratory of Theoretical Physics, 
Joint Institute for Nuclear Research, 141980 Dubna, Russia}

\maketitle

\begin{abstract}
While the dynamics for three-dimensional axially symmetric 
two-electron quantum dots with parabolic confinement potentials
is in general  non-separable we  have found an exact separability
with three quantum numbers for specific values of the magnetic field. 
Furthermore, it is shown that the magnetic properties such as the
magnetic moment and the susceptibility are sensitive 
to the presence and strength of a vertical confinement.
Using a semiclassical approach  the calculation of
the eigenvalues reduces to simple quadratures providing a transparent and
almost analytical quantization of the quantum dot energy levels which
differ from the exact energies only by a few percent.
\end{abstract}

\vskip 0.3cm


\vskip 1cm

Current nanofabrication technology allows one to control the size and
 shape of quantum dots \cite{Tur,Bim,Tap}. Due to the confinement 
of the electrons in all three
spatial directions the energy spectrum is quantized creating
excellent experimental and theoretical opportunities
to study  {\it controlled} single-particle and collective 
dynamics at the atomic scale.
For example, depending on the experimental setup, the spectrum of 
a quantum dot displays shell structure \cite{Kouw,Mac} or follows
predictions of random matrix theory (for a review see \cite{Al}).
Furthermore, it becomes possible to trace the transition
from a quantum mechanical  to an almost classical regime.

Few-electron quantum dots have attracted special attention \cite{Kouw,Mak}, 
since they may provide a natural realization of a quantum bit \cite{Tur}.
The simplest quantum dot with the essential features of more complex 
systems contains two electrons.
Experimental data, including
transport measurements \cite{Su} and spin oscillations in 
the ground state under a perpendicular magnetic field \cite{Ash}, 
have been explained quantum mechanically as a result of the interplay 
between the two-dimensional lateral confinement potential, 
electron correlations and 
the magnetic field \cite{Mer,Din}.  
While in these experiments effects of the 
third spatial dimension are somewhat hidden, they naturally come into play 
with a tilted magnetic field \cite{MHP}. However, even with a perpendicular
magnetic field the vertical confinement has at least 
two important consequences which will be worked
out in the following:
 first, it changes the magnetic moment and
susceptibility with respect to the 2D results, second, the generically 
non-separable 3D dynamics
becomes separable for certain values of 
the magnetic field. We will use a
semiclassical description 
which offers a simple and accurate approach to 
explore the effects of dimensionality in  quantum dots.
In contrast to a circular (2D) two-electron quantum dot 
whose classical dynamics is always separable and therefore regular, the corresponding 
3D-system with axial symmetry is in general a 
non-integrable problem with typical features of mixed dynamics
(regular/chaotic).

The Hamiltonian for the 3D two-electron quantum dot reads

\beq
\label{ham}
H = \sum_{j=1}^2{\Bigg\{} \frac{1}{2m^*}
({\bf p}_j \!-\!\!\frac{e}{c}  {\bf A}_j)^2+\frac{m^*}{2}\left[
\omega _0^2(x_{j}^2\!+\!y_{j}^2)+\omega _z^2z_{j}^2 \right]\! {\Bigg\}}
+ V_C + H_{\rm spin},
\eeq
where $V_C = \alpha/{|{\bf r}_1 - {\bf r}_2|}$ is the Coulomb energy 
($\alpha = e^2/(4\pi \varepsilon \varepsilon_0)$)
and 
$H_{\rm spin} = g^*({\bf s}_1 + {\bf s}_2) \!\cdot\! {\bf B}$ describes 
the Zeeman energy. 
Here $m^*$ and  $ g^*$ are the effective electron mass and 
$g$-factor,  respectively, and $\varepsilon$ is the dielectric constant.
The confining potential is approximated with a 3D axially-symmetric
harmonic oscillator 
and $\hbar \omega_z \neq \hbar \omega_0$ are the  energy scales of
confinement  
in the $z$-direction and in the $xy$-plane, respectively.
For the typical voltage $\sim 1$ V applied to the gate, the
confining potential is  some eV deep which is large compared 
to  the few meV of the confining frequency \cite{Bim,Tap}. 
Hence,  the electron
wave function is localized close to the minimum of the well which 
always can be approximated by a parabolic potential.
In real samples the electron-electron interaction is usually 
screened. However, the pure Coulomb interaction should suffice
to understand the main features of the system.
For the perpendicular magnetic field
$({\bf B}\parallel z)$ we choose a gauge described by the
vector ${\bf A} =[{\bf B}\times {\bf  r}]/2 = \frac{1}{2}B(-y, x,0)$.
Introducing the relative and center-of-mass coordinates
${\bf r} = {\bf r}_1 - {\bf r}_2$, $ {\bf R} = 
\frac{1}{2}({\bf r}_1+{\bf r}_2)$, 
the Hamiltonian, Eq.(\ref{ham}), can be separated into the
center-of-mass (CM) $H_{\rm CM}$  and relative-motion (RM) 
$H_{\rm rel}$ terms: $H=H_{\rm CM}+H_{\rm rel}+H_{\rm spin}$.
The solution to the CM-Hamiltonian is well known \cite{Fock} and 
the effect of the Zeeman energy has been discussed in \cite{Mer,Din}.
In the following we will concentrate on the dynamics of $H_{\rm rel}$.

For our analysis it is convenient to use cylindrical {\it scaled}
coordinates, $\tilde\rho = \rho/l_0$, ${\tilde p}_{\rho} =
p_{\rho}l_0/\hbar$, $\tilde z = z/l_0$, $ {\tilde p}_{z} = p_z
l_0/\hbar$, where $l_{0}=(\hbar/\mu\omega_0)^{1/2}$ is the
characteristic length of the confinement potential with the reduced
mass $\mu = m^*/2$.  The strength parameter $\alpha$ of the Coulomb
repulsion goes over to $\lambda = 2\alpha/(\hbar \omega_0 l_0)$. 
Using the effective mass $m^*=0.067 m_e$, the dielectric constant
$\varepsilon=12$, which are typical for GaAs, and the confining
frequency $\hbar \omega_0$=3 meV, we obtain $\lambda \approx 3 $.
Hereafter, for the sake of simplicity, we drop
the tilde, i.e. for the scaled variables we use the same symbols as
before scaling. 

In these variables the Hamiltonian for the relative motion  
takes a particular simple form (in units of $\hbar \omega_0$) 
\beq
\epsilon \equiv \frac{H_{\rm rel}}{\hbar\omega_0} = \frac{1}{2}
\left[ p_\rho^2 + \frac{m^2}{\rho^2} + p_z^2 +
\left(\frac{\omega_\rho}{\omega_0}\right)^{\!\!2}\!\rho^2 +
\left(\frac{\omega_z}{\omega_0}\right)^{\!\!2}\!z^2
+ \frac{\lambda}{\sqrt{\rho^2+z^2}} \right] - \frac{\omega_L}{\omega_0}\,m,
\label{relham}
\eeq
where $m = l_z/\hbar$, $\omega_L=eB/2m^*\!c$
is the Larmor frequency and 
\begin{equation}
\label{frequ}
\omega_{\rho}=(\omega_{L}^{2}+\omega_{0}^{2})^{1/2}
\end{equation}
is the effective confinement frequency in the $\rho$-coordinate which
depends through $\omega_{L}$ on the magnetic field.

Due to the cylindrical symmetry, the $z$-component $l_z \equiv p_\phi$
of the angular momentum is conserved and the motion in $\phi$ is
separated from the motion in the $(\rho, z)$-plane.  Since the Coulomb
term couples the two coordinates, the problem is in
general non-integrable which is reflected in the Poincar\'e sections
shown in Fig.1 for increasing magnetic field.  The chosen 
ratio $\omega_z/\omega_0 = 3$ is of the same
order of magnitude as in the experiment \cite{MHP}.  For $\omega_L
= 0$ and small values of $m$ the motion is mainly chaotic (see Fig.1a). 
 With the magnetic field the frequency of oscillations
along the $\rho$-coordinate can be controlled which leads to
qualitatively different dynamical situations (Fig.1b-d). 
For equal effective confinement frequencies
$\omega_\rho^2=\omega_z^2$, the Hamiltonian Eq.(\ref{relham})
becomes separable in spherical coordinates and the dynamics is
integrable (Fig.1c).
For two other limiting cases, the dynamics is nearly integrable, 
namely in the limit $m\to \infty$ and for $\omega_{z}\to \infty$. The 
latter case represents a two-dimensional quantum dot,
classically, we have $p_{z}, z\to 0$ in this limit. 

The semiclassical quantization of the  {\it circular  2D quantum 
dot} is particularly simple since it reduces to a 1D WKB quantization 
of the $\rho$-motion due to the separability of the problem. For given $m$ and $p_{z}=z=0$ 
the momentum $p_{\rho}$ determined from Eq.(\ref{relham}) enters the action 
integral 
\beq
I_{\rho} = \frac{\hbar}{2\pi}\oint p_{\rho}\,d\rho = 
\frac{\hbar}{\pi}\int_{\rho_{\rm min}}^{\rho_{\rm max}} |p_\rho|\,d\rho,
\label{action}
\eeq
with the turning points $\rho_{\rm min}$, $\rho_{\rm max}$ as 
the positive roots of
equation $p_\rho(\rho) = 0$. The WKB quantization conditions
\beq 
I_{\rho}(\epsilon) = \hbar\,(n_{\rho} + 
\hbox{$\frac{1}{2}$}), \quad n_\rho = 0,1,..., \quad m = 0,\pm 1,... 
\eeq
determine the energy levels. For non-interacting electrons ($\lambda = 0$) the
analytical calculation of the action integral leads to the 
(quantum mechanically exact) eigen-energies 
\beq
\epsilon = \sqrt{1+\left(\frac{\omega_L}{\omega_0}\right)^{\!\!2}}\,
(2 n_\rho\! + \vert\,m\,\vert + 1) - \frac{\omega_L}{\omega_0}\,m,
\eeq
which are the well known Fock-Darwin energies \cite{Fock}.
For $\lambda \neq 0$,  we calculate the action 
integral Eq.(\ref{action}) numerically with a
few iterations to determine the quantum eigenvalues.

The energy spectra for non-interacting and interacting
electrons are shown in Fig.2. In the interacting 
case the semiclassical result, although not exact (the error is less
than $1\%$),  
reproduces very well the quantum mechanical results
 \cite{Din,Mer}.

Turning now to 
the  {3D quantum dot} we have seen that the dynamics is separable
for  
$\omega_z^2 = \omega_{\rho}^2 \equiv{\omega_L^*}^2+\omega_{0}^2 $
and the Hamiltonian Eq.~(\ref{relham}) in scaled spherical coordinates
takes the form 
\beq
\epsilon = \frac{1}{2} \left\{ 
p_r^2 + \left(\frac{\omega_z}{\omega_0}\right)^{\!\!2}\!r^2 + 
\frac{\lambda}{r} + \frac{({\bf l}/\hbar)^2}{r^2} \right\} -
\frac{\omega_L^*}{\omega_0}\,m\,.
\label{relsp}
\eeq
In this case the square of the total angular momentum ${\bf l}^2$ is
an additional integral of motion.  Therefore, the classical dynamics
reduces again to a one-dimensional, radial problem.  Using 
 Eq.(\ref{relsp}) and calculating the action integral for the
radial motion analogous to that in Eq.(\ref{action}) (i.e. with $r$
instead of $\rho$), we obtain the energy levels from the standard WKB
quantization conditions
\bea 
&&I_r(\epsilon) = \hbar\,(n_r + \hbox{$\frac{1}{2}$}), \quad  
\vert\,{\bf l}\,\vert = \hbar\,(l + \hbox{$\frac{1}{2}$}), \nonumber\\
&&n_r, l = 0, 1, ..., \quad m = 0, \pm 1, ..., \pm l.
\eea
Note that it is only the magnetic field which generates the spherical
symmetry of the problem and therefore its separability leading to
three good quantum numbers $n_r$, $l$ and $m$.
 
In the general case of an {\it axially symmetric 3D quantum dot} we
have non-integrable motion and a semiclassical quantization is neither
straight forward nor does it give results which allow for a simple
understanding of the dynamics.  Therefore, we make use of the fact
that in real samples the confining potential in the $z$-direction is
much stronger than in the $xy$-plane which allows us to analyze the 3D
non-integrable system with the
'removal of resonances' method (RRM) \cite{LL}.  To
lowest order the RRM consists of averaging the Hamilton function over
the fastest angle of the unperturbed motion $(\lambda=0)$ after
rewriting coordinates  and  momenta in terms of action-angle
variables $(J_{\rho}, J_z, \theta_{\rho}, \theta_z)$: 
\bea
&&\!\!\!\!\rho^2 = \frac{\omega_0}{\omega_{\rho}}\left(2j_\rho+|m|-
2\sqrt{j_{\rho}(j_{\rho}+|m|)}\cos 2\theta_{\rho}\right)\quad \quad, \\
&&\!\!\!\!z^2  = \frac{2j_z\omega_0}{\omega_z}\,\sin^2\!\theta_z \quad,  
\eea
and $p_{\rho}={\dot{\rho}}$, $p_z={\dot{z}}$. 
Here, $j_z=J_z/\hbar$ and $j_{\rho}=J_{\rho}/\hbar$. 
If $\omega_z > \omega_\rho$ one averages over the angle 
$\theta_z = \omega_zt$.  As a result, the motion effectively 
decouples into an unperturbed motion in the $z$-coordinate governed 
by the potential
$(\omega_z / \omega_0)^2z^2/2$ and into the relative motion in the
$\rho$-coordinate governed by the effective potential
\beq
V_{\rm eff}(\rho, j_z) =  
\frac{1}{2}\!\left(\frac{\omega_\rho}{\omega_{0}}\right)^{\!\!2}\!\rho^2
+ \frac{m^2}{2\rho^2} 
+ \frac{\lambda}{\pi\rho}\,K\!\left(\!-2\frac{\omega_{0}}{\omega_z}
\frac{j_z}{\rho^2}\right),
\label{effham}
\eeq
where $K$ is the first elliptic integral. 
Hence, the effective Hamiltonian reads
\beq
\epsilon = \frac{p_\rho^2}{2}  + V_{\rm eff} - 
\frac{\omega_L}{\omega_0}\,m
+ \frac{\omega_z}{\omega_0}\,j_z.
\label{scaleffen}
\eeq
Applying a similar procedure as in the 2D case, we calculate the
action integral numerically.  The momentum $p_\rho$ is determined from
Eq.(\ref{scaleffen}) and the turning points $\rho_{\rm min}$, 
$\rho_{\rm max}$
are as usual the (positive) roots of the equation $p_\rho(\rho) = 0$. 
Finally, the WKB-quantization conditions
\bea
&& I_{\rho}(\epsilon) = \hbar\,(n_{\rho} +  \hbox{$\frac{1}{2}$}), \quad 
j_z = n_z + \hbox{$\frac{1}{2}$},\nonumber\\ 
&& n_\rho, n_z = 0,1,2,..., \quad m = 0,\pm 1, \pm 2,...,
\eea
determine the energy levels.

Comparing the exact results for eigen-energies for the spherical case
$\omega_z/\omega_\rho = 1$
we found good agreement even for large
values of the magnetic field (Fig.3a) although RRM is expected
to work best for $\omega_\rho/\omega_z <1$. Without magnetic field we
have $\omega_\rho/\omega_z =1/3$ which means that the motion in $z$ and $\rho$
approximately decouples justifying the widely used 2D approximation. This
is also reflected in the small difference between 2D and 3D results
(compare Fig.2b with Fig.3a at $\omega_{L}=0$).
Turning on the magnetic field increases the coupling of the dynamics in 
$\rho$ and $z$  which allows the two electrons eventually to access
the full 3D space. As a consequence, the electrons can avoid each other
more effectively and the Coulomb interaction has a smaller effect on the
3D spectrum than on the 2D spectrum which is most clearly visible for the
$m=0$ energies, see Fig.3a. We can understand this effect
quantitatively by averaging the elliptic integral in 
Eq.(\ref{effham}) over the unperturbed ($\lambda =0$) motion in $\rho$.
It gives rise to an effective charge in the Coulomb interaction
$V_C\approx{\lambda_{\mathrm{ eff}}}/2{\rho}$, where
\beq
\lambda_{\mathrm{ eff}}=\frac{2\lambda}{\pi^2}\int_0^\pi\!\!
K\!\left(\!-\frac{\omega_{\rho}/\omega_z}
{1+|m|-{\sqrt{1+2|m|}}\cos{2\theta_\rho}}\right)\!d\theta_{\rho}
\label{efk}
\eeq
for $n_{\rho}=n_z=0\quad (j_ \rho=j_z=1/2)$.
The 3D energy quantized with this effective charge for the repulsion
is close to the full interaction (dotted line in Fig.3a).

The effective charge $\lambda_{\mathrm{ eff}}/\lambda$ as a function
of $\omega_\rho/\omega_z$ for different $m$
is shown in Fig.3b. 
The maximum repulsion at $\omega_\rho/\omega_z= 0$ corresponds
with $\omega_z\to\infty$ to the 2D case. The 3D case without magnetic
field starts for our parameters $\omega_\rho/\omega_z= 1/3$ at some
value $\lambda_{\mathrm{ eff}}/\lambda <1$ which decreases 
further for increasing
$\omega_\rho/\omega_z$, i.e., increasing magnetic field. 
This explains quantitatively through the effective charge the difference
of the effect of a magnetic field  on a quantum spectrum in 2D and 3D cases. 
However, this difference becomes weaker for larger $m$ as it is seen in
Fig.3b.

Although
the ground state as a function of the magnetic field is formed piecewise
by levels of {\it increasing} $m$ and alternating singlet-triplet character
(see Fig.3a) the magnetic properties of the ground state nevertheless
reveal the dimensional difference between 2D and 3D.
At temperature $T=0$ the dot is
in the ground state and the magnetic moment and the magnetic 
susceptibility are  defined by
$\mu_{\rm mag} = -\partial E_{\rm gr}/\partial B$
and $\chi  = \partial \mu_{\rm mag} /\partial B$, respectively.
Both quantities exhibit discontinuities as a function of the magnetic 
field due to the symmetry
changes (with respect to $m$ and spin).
We find that these discontinuities shift 
when going from the 2D quantum dot to the 3D case as shown 
in Fig.4.

By relaxing the restriction of two dimensions for a quantum dot and
working in the physical three-dimensional space we have investigated
physical examples of non-integrable systems close to integrability. 
For this situation the RRM method is naturally justified, since the
confining frequencies in quantum dots obey the condition $\hbar
\omega_z \gg \hbar\omega_0$.  The WKB-approach provides a simple and
transparent way to calculate the  spectrum of the 3D
two-electron quantum dot even for {\it small} values of the quantum
numbers. We have found 
that at {\it specific} values of the magnetic field 
$\omega_L^*=\sqrt{\omega_{z}^2 - \omega_{0}^2}$
an axially-symmetric quantum dot exhibits spherical symmetry and its
dynamics becomes completely separable with three integrals of motion
and three corresponding quantum numbers.  We have shown that the
confinement in the $z$-direction, neglected in the 2D description of
quantum dots, does have an influence on the spectrum and consequently
also on magnetic properties of the dot.  In fact, by changing the
confining frequency in the $z$-direction only slightly one can
increase or decrease the magnetic moment and the susceptibility, i.e.
one can control the magnetic properties of the two-electron quantum
dot.

\vskip 1cm

{\bf Figure captions:}
\vskip 0.2cm

{\bf Fig.1.} Poincar\'e surfaces of sections $z = 0$, $p_z > 0$ 
of the relative motion for
the axially-symmetric 3D two-electron quantum dot ($\omega_z/\omega_0 = 3$,
$\lambda = 3$, $m = 0$, $\epsilon = 5$) in the 
magnetic field for: (a) $\omega_L =
0$, (b) $\omega_L/\omega_0 = 2.5$, (c) $\omega_L/\omega_0 = \sqrt{8}$ and (d)
$\omega_L/\omega_0 = 3.3$. The section (c) indicates that for the corresponding
value of the magnetic field the system is integrable.

{\bf Fig.2.} The energy spectrum of a circular 2D quantum dot (in units
$\hbar\omega_0$) as a function of the ratio $\omega_L/\omega_0$ for
$n_\rho=0$ and $m = 0,...,9$ in the cases: (a) $\lambda=0$ and 
(b) $\lambda=3$.

{\bf Fig.3.} (a) The comparison between energy levels (in units
$\hbar\omega_0$) of the axially-symmetric 3D quantum dot with
$\omega_z/\omega_0 = 3$ and $\lambda=3$ for $n_\rho=n_z = 0$
and $m = 0,...,9$ obtained using the RRM (full lines) and
exact results for the spherical case (circles).
The inset shows a good agreement between the RRM and the exact
results. The dashed and dotted lines display the energy level with $m=0$ 
for the 2D and 3D cases with 
$\lambda_{\mathrm{ eff}}$ 
at $\omega_0/\omega_z=0$ and $1/3$, respectively.
(b) The dependence of the effective strength
of the Coulomb interaction $\lambda_{\mathrm{ eff}}/\lambda$ 
on the ratio $\omega_{\rho}/\omega_z$.

{\bf Fig.4.} Magnetic moments $\mu_{\rm mag}$ 
(a) in the units of effective Bohr
magneton $\mu_B^* = (m_e/m^*)\,\mu_B$ 
and the magnetic susceptibility $\chi$ (b)
for the 2D (dashed lines) and 3D (full lines) cases 
as a function of  the  magnetic field strength (in $\omega_L/\omega_0$-units). 
We use the same parameters as in Figs.2,3.

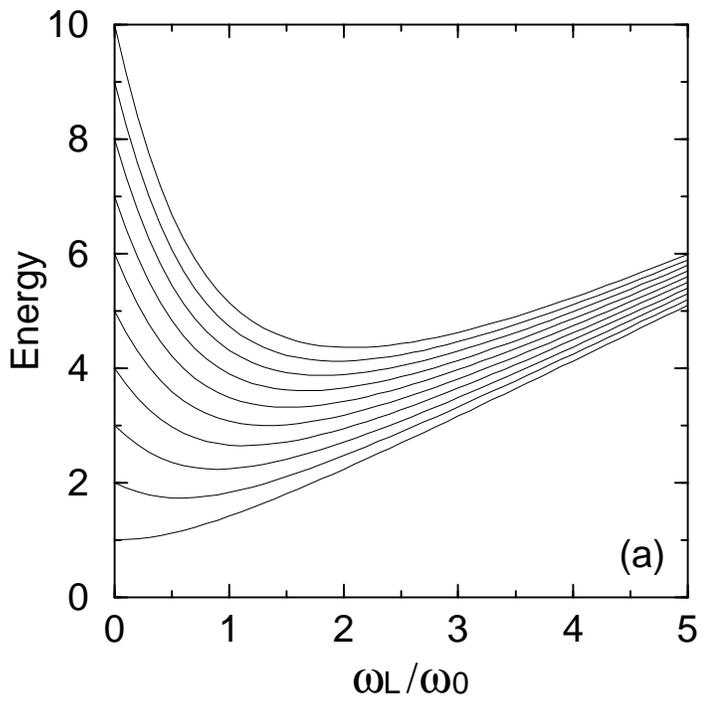
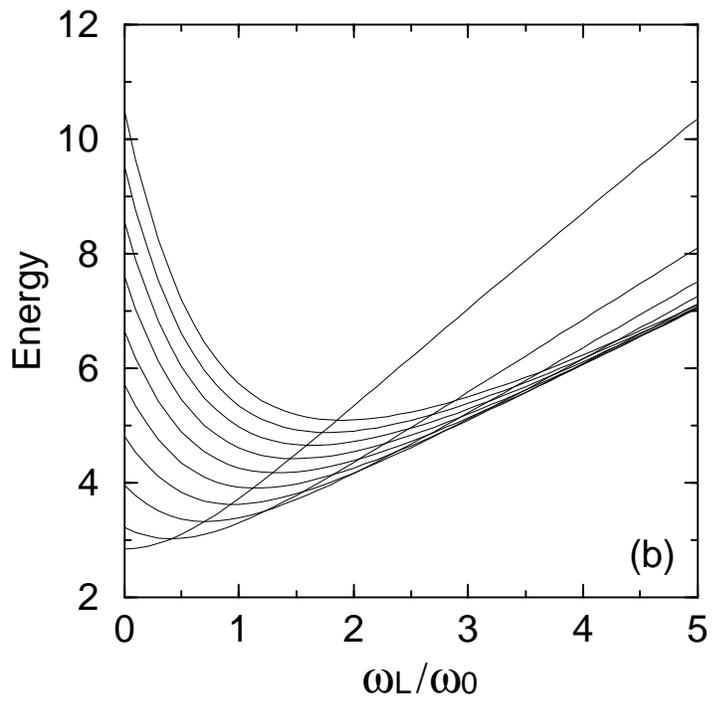
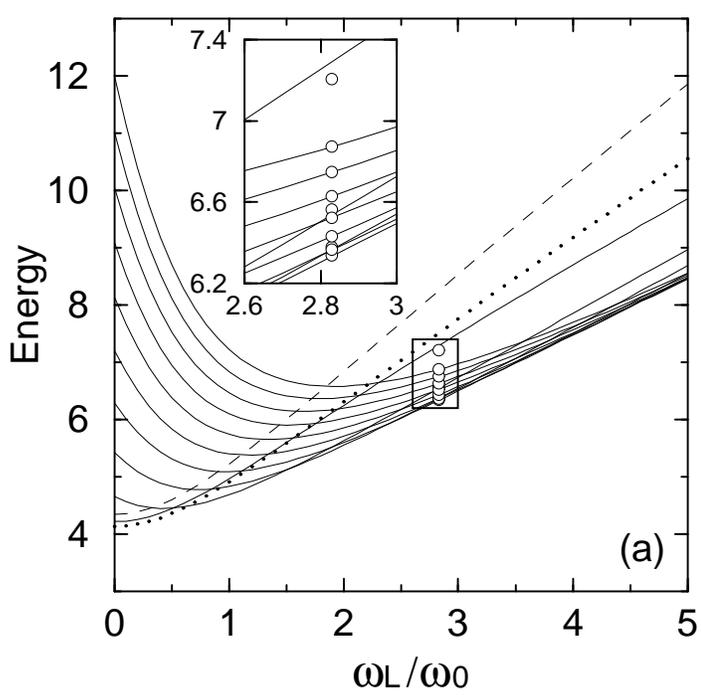
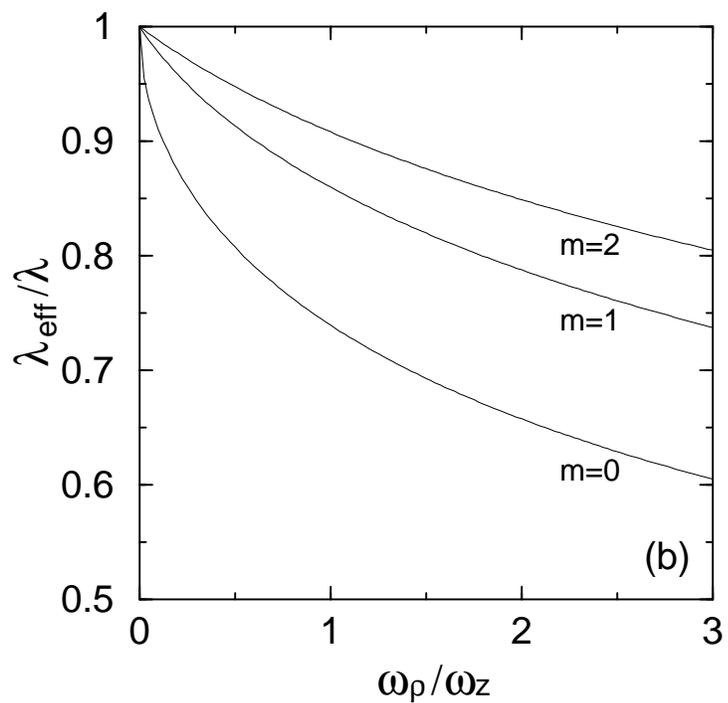
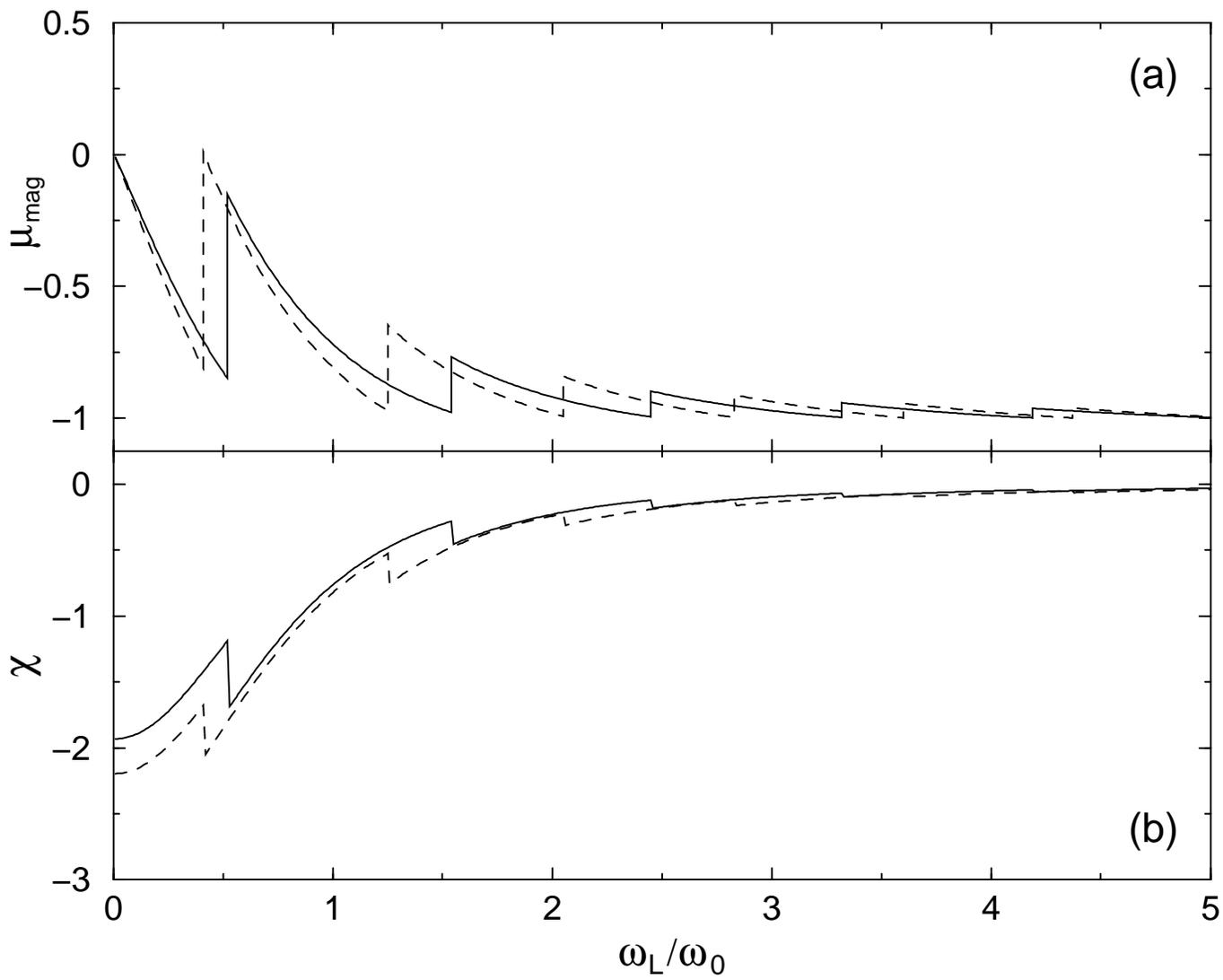
\end{document}